# Effect of Structural Transition on Magnetic and Dielectric Properties of La and Mn co-substituted $BiFeO_3$ Ceramics

Pawan Kumar and Manoranjan Kar*

Department of Physics, Indian Institute of Technology Patna, Patna-800013, India.

* Corresponding author Email: mano@iitp.ac.in, Ph: +91612-2552013, Fax: +91612- 2277383

**Abstract:**

$Bi_{1-x}La_xFe_{1-x}Mn_xO_3$ (x = 0.000 - 0.300) ceramics prepared by the tartaric acid modified sol-gel technique have been studied to analyze the effect of composition driven structural transition on the magnetic properties of bismuth ferrite ($BiFeO_3$). It was found that the co-substitution of La & Mn at Bi & Fe sites leads to suppression of impurity phases ($Bi_2Fe_4O_9$ and $Bi_{25}FeO_{40}$) which generally appear in $BiFeO_3$. The quantitative crystallographic phase analysis has been carried out by double phase Rietveld analysis of all the XRD patterns which indicates the existence of compositional driven crystal structure transformation from rhombohederal (space group *R3c*) to the orthorhombic (space group *Pbnm*) with the increase in substitution concentration. The 5% co-substituted sample exhibit high remnant magnetization i.e. about 15 times that of $BiFeO_3$ which is due to the suppression of cycloid spin structure and enhanced canting angle of antiferromagnetically ordered spins caused by the crystal lattice distortion. However, further increase in substitution results in the reduction of remnant magnetization and coercivity due to the appearance of complete antiferromagnetic ordering in the orthorhombic structure because of the significant contribution from the crystallographic phase of *Pbnm* space group (as obtained from double phase Rietveld analysis). The frequency independent higher dielectric constant and lower dielectric loss were observed for 5% co-substitution. Hence, this particular composition may be interesting for device applications.

## 1. Introduction

Magnetoelectric Multiferroic materials showing both electric as well as magnetic ordering within a single crystallographic phase are attracting much attention due to their promising technological applications, such as multistate memory devices, spintronic devices, magnetically modulated transducers, ultrafast optoelectronic devices and sensors [1-4]. However, these kinds of materials are rare because ferroelectric ordering in $ABO_3$ perovskite needs transition metal ions with $d^0$ electronic configuration, whereas magnetic ordering needs transition metals with partially filled d orbitals, but these two conditions within single crystal phase are mutually exclusive. Among all the identified single-phase magnetoelectric materials, $BiFeO_3$ (BFO) is most promising one because it shows both electric and G-type canted antiferromagnetic ordering well above room temperature ($T_c$ = 1103 K and $T_N$ = 643 K) [5]. The crystal structure for polar phase of BFO (at room temperature) is rhombohederally distorted perovskite structure with *R3c* space group. The *R3c* symmetry allows the existence of weak ferromagnetic moment due to Dzyaloshinky-Moriya interaction but the cycloid spin structure with the periodicity of ~62 nm prevents net magnetization which leads to net zero magnetization [6-9]. However, BFO is very difficult to be prepared in phase pure ceramics or thin film because of its narrow temperature range of phase stabilization. It results in oxygen vacancies, multiple valence state of Fe ions and synthesis of impurity phase as $Bi_2Fe_4O_9$ and $Bi_{25}FeO_{40}$ which increases the leakage currents and make it unsuitable for the device applications. From the literature survey, it has been observed that the A-site and/or B-site substitution in BFO has been reported as the most effective way to reduce the impurity phases and enhance magnetoelectric coupling constant by creating the lattice strain due the ionic size mismatch between host and substituting cations [10-15]. Several attempts have also been made to prepare phase pure BFO by chemical route and the solid state routes followed by leaching with nitric acid normally used to eliminate above impurity phases which leads to the formation of coarser powders and its poor reproducibility [16, 17]. Hence, we have adopted the chemical route of synthesis for uniform particle size and better reproducibility.

The cycloid suppression in Lanthanum (La) substituted BFO ceramics has been shown using nuclear magnetic resonance (NMR) measurements which was correlated with the structural transition from rhombohedral to orthorhombic crystal structure [10, 14, 15]. Weak ferromagnetic ordering induced due to La substitution at A-site in BFO has been observed by Y. Du et al [18]. The Mn substitution at Fe site of BFO is reported to inhibit the grain growth which resulted in reduced particle size [19] and improve the magnetic as well as electric properties [20, 21]. S. Chauhan et al. has reported the structural phase transition in 15% Mn doped BFO ceramics due to the distortion in the rhombohedral structure with increasing Mn substitution which resulted in significant enhancement in magnetization [22]. B. Li et al have reported the weak ferromagnetism in $Bi_{0.9}La_{0.1}Fe_{0.95}Mn_{0.05}O_3$ nanofibers which exhibited a larger $M_s$ value owing to the canting of antiferromagnetically ordered spins and an increase in ferromagnetic ordering [23]. A. Lahmar et al have reported the structural transition and improved dielectric as well as ferroelectricity in $BiFeO_3$–$LaMnO_3$ solid solution thin films grown on Pt-terminated silicon substrates up to 10% substitution concentration [24] which induces structural changes to orthorhombic structures.

However, a detailed study of the influence of crystal structural transition on physical properties is yet to be explained on La and Mn co-substituted BFO ceramics. Hence, we have undertaken a study on La and Mn co-substituted BFO prepared by the tartaric acid modified sol-gel method and carry out double phase Rietveld analysis to study the crystallographic phases and its correlation with the physical properties.

**2. Methods**

$Bi_{1-x}La_xFe_{1-x}Mn_xO_3$ with x = 0.000, 0.015, 0.025, 0.050, 0.100, 0.150, 0.200, 0.250 and 0.300 (named as BFO, BLFM-015, BLFM-025, BLFM-05, BLFM-10, BLFM-15, BLFM-20, BLFM-25 and BLFM-30 respectively) were prepared by a tartaric acid modified sol-gel technique [25]. Here, the starting materials as bismuth nitrate, iron nitrate, lanthanum acetate, manganese acetate and tartaric acid (purity ≥ 99.0%) were carefully weighted in stoichiometric proportion. The molar ratio of metal nitrates to tartaric acid of 1:2 was taken because there were two cations in BFO. The tartaric acid plays an important role in the process of forming pure and co-substituted $BiFeO_3$ nanoparticles and carboxyl in tartaric acid makes bonds with all metal cations present in the solvent. Hence, the obtained powder exists as amorphous precursors before calcinations. When they are heated, a driving force for the nucleation is related to reduction of free energy due to transformation from the metastable amorphous to the crystalline state in equilibrium and tartaric acid vaporizes in the annealing process and the M-O bonds are broken following by the nucleation and crystal growth of small and uniform nanoparticles [26]. The resulting material was thoroughly grinded and annealed at 700 $^oC$ for 3 hours.

The crystallographic phases of all the ceramics were studied by the powder X-ray diffraction (XRD) using 18 kW Cu-rotating anode based Rigaku TTRX III diffractometer, Japan) with CuKα radiation (λ = 1.5418 Å) operating in the Bragg-Brentano geometry in a 2θ range of $10^o$-$120^o$ at a scan step of $0.01^o$ (counting time for each step was 2.0 Sec). The microstructural properties and elemental analysis were investigated by using Field Emission Scanning Electron Microscopy (FE-SEM) (Hitachi S-4800, JAPAN) operating at an accelerating voltage of 10 kV and equipped with energy-dispersive X-ray spectroscopic capability. Room temperature Raman spectra were studied in the backscattering geometry using confocal micro-Raman spectrometer (Seki Technotron Corp., Japan) with the 514.5 nm laser line as excitation source by STR 750 RAMAN Spectrograph using a 100 X microscope. Fourier Transform Infrared Spectra (FT-IR) were recorded at room temperature using the Perkin Elmer (model 400) in the range from 350 to 1200 $cm^{-1}$. The diffuse reflectance spectroscopy has been carried out by using LAMBDA 35 UV-visible spectrophotometer in the range from 200 to 1100 nm. Room temperature magnetization – magnetic field (M-H) measurement for all the ceramics were carried out by magnetic properties measurement system (MPMS), Quantum Design Inc., USA in the applied magnetic field of maximum ± 80 kOe. The parallel plane surfaces of pellets were coated with the silver paste for dielectric measurement. The dielectric constant and loss measurement were performed using the impedance analyzer (N4L PSM 1735) in the frequency range of 500 Hz – 5 MHz at the room temperature (30 $^{o}$C).

All XRD patterns were analyzed employing Rietveld refinement technique with the help of Fullprof suite program [27]. The patterns for all the ceramics could be refined using the *R3c* and/or *Pbnm* space groups. The parameters such as, Zero correction, scale factor, half width parameters, lattice parameters, atomic fractional position coordinates, thermal parameters were varied during refinement process. Background and peak shape were refined by the sixth order polynomial and pseudo-voigt function respectively.

**3. Results and discussion**

Figure 1 shows the XRD patterns of the present ceramics. It is important to note that small amount of $Bi_2Fe_4O_9$ and $Bi_{25}FeO_{40}$ impurities were observed along with the BFO phase for co-substituted ceramics with co-substitutions up to 2.5%. It indicates that the co-substitution of La & Mn has suppressed the impurity phase formation in the BLFM ceramics which is the main concern for single phase BFO because Bi and/or Fe vacancies lead to formation of these impurity phases [28]. The distinct peaks corresponding to the planes (104) and (110); (006) and (202); (116) and (122); (018) and (300) around 32$^{o}$, 39$^{o}$, 51$^{o}$ and 57$^{o}$ respectively merge into single peak with the increase in the La and Mn concentration, which is clearly visible higher co-substitutions (≥ 5%). Also, the intensity of a peak corresponding to the planes (006) at around 39$^o$ which originates from the rhombohedral structure became weaker with the increase in co-substitution. These results indicate the compositional driven crystallographic phase transition from rhombohedral to orthorhombic which is consistent with the Rietveld analysis as discussed below. The Goldschmidt tolerance factor (t) is used to quantify the structural stability of $ABO_3$ ($BiFeO_3$) structure (perovskites) which is defined as,

$$t = \frac{(< r_A > + r_o)}{\sqrt{2}(< r_B > + r_o)} \quad \text{-------------- (1)}$$

The $<r_A>$ and $<r_B>$ in equation 1 are the average radius of A site and B site cations respectively and $r_o$ is the ionic radius of oxygen. When the value of "t" is smaller than one (mentioned in Table 1), the compressive strain acts on the Fe-O bonds and hence on Bi-O bonds which infers that the oxygen octahedral must buckle in order to have minimum lattice strain. This induced lattice distortion suppresses the rhombohedral phase and leads to the evolution of modified crystallographic phases. The particle size has been calculated using Scherrer's formula [29] which is defined as,

$$D = \frac{k\lambda}{\beta Cos\theta} \quad \text{-------------- (2)}$$

Where constant $k$ depends upon the shape of the crystallite size (= 0.89, assuming the spherical particles), $\beta$ = Full width at Half Maximum (FWHM) of Intensity (a.u.) vs. 2θ profile, $\lambda$ is the wavelength of the $Cu\ K_\alpha$ radiation (= 0.1542 nm), θ is the Bragg's diffraction angle and $D$ is the crystallite size. Here, the value of FWHM has been used from Rietveld refinement of corresponding XRD patterns. The observed FWHM has been corrected by considering the instrumental broadening ($\beta = \sqrt{\beta_{obs}^2 - \beta_{ins}^2}$ , $\beta_{obs}$= Observed FWHM and $\beta_{ins} = Instrumental\ FWHM$). According to Rietveld method [30], the individual contributions to the broadening of reflections can be expressed as,

$$FWHM^2 = (U + D_{ST}^2)(\tan^2\theta) + V(\tan\theta) + W + \frac{IG}{Cos^2\theta} \quad \text{-------------- (3)}$$

Where $U$, $V$ and $W$ are the usual peak shape parameters, $IG$ is a measure of the isotropic size effect, $D_{ST}$ =coefficient related to strain. As shown in table 1, the crystallite size decreases with the increase of substituent ion concentration up to co-substitution of 15%. This implies the development of lattice strain inside the lattice due to ionic size mismatch between host and substituent cations which lead to local structural disorder and reduces the rate of nucleation resulting in decrease of crystallite size. However, the crystallite size increases when substitution concentration increases beyond 15 % of co-substitution due to decrease in overall lattice strain decreases. The lattice constants and corresponding unit cell volume decreases with the increase in the substituent concentration due to smaller radius of substituents than that of host cations, which can be inferred from the shift of the XRD peak towards higher 2θ value as shown in Fig. 2. However, their values increases when co-substitution concentration increases beyond 15 % (near MPB) because the orthorhombic phase becomes dominant and overall lattice strain decrease. The gradual shift in XRD peak indicates that the substituents get substituted in the BFO lattice.

The rhombohedral phase was considered for the Rietveld refinement of BFO XRD pattern because it has the characteristic doublet of highest intensity peaks but it merges for the co-substituted BFO which suggests the presence of modified crystallographic symmetry. Rigorous fitting with the different structural models (*R*3*c*, *R*3*c* + *Pbnm*, *R*3*c* + *P*4*mm*, *R*3*c* + *Pm*3*m*, *R*-3*c* + *Pbnm*, etc.) showed that the observed XRD patterns of co-substituted ceramics are a result of the superposition of two spectral contributions (*R3c* + *Pbnm*). It is worth noting that we have not got satisfactory results by taking a single space group in Rietveld refinement of XRD patterns of co-substituted ceramics. The typical Rietveld refined XRD patterns for BLFM-05 and BLFM-10 has been shown in Figs. 3 (a) and (b) respectively. The fractional atomic coordinates for the Rietveld refinement of XRD patterns of BLFM-05 and respective contribution of both crystallographic phases has been given in Table 2 and 3 respectively. The crystallographic phase contribution of *Pbnm* space group increases with the increase in the co-substitution percentage.

Typical FE-SEM images are shown in figures 4 (a) and (b) for BLFM-025 and BLFM-05 respectively. Average crystallite sizes are found to be 52 and 45 nm (close to the values obtained from XRD patterns) for BLFM-025 and BLFM-05 respectively. The elemental analysis for all the samples were carried out using the EDS analysis. Typical EDS pattern of ceramics BLFM-025 and BLFM-05 have been shown in Figs. 4 (c) and (d). The EDS analysis suggests the presence of Bi, Fe, La, Mn and O elements in the co-substituted ceramics and it showed the atomic ratio of A site (Bi/La) and B site (Fe/Mn) cations is approximately 1:1. No extra peaks have been traced which indicate that there is no contamination in the ceramics. However, ratio of A site/B site and O site ions is found to be around 1:2.94 instead of 1:3. It suggests the presence of oxygen vacancies in the sample which has been reflected in our dielectric measurement as discussed in the later section.

Since Raman Spectroscopy is sensitive to the atomic displacements, we have carried out Raman spectroscopy in order to further investigate the structural transition. Eighteen (18) optical phonon modes: $4A_1+5A_2+9E$ for $BiFeO_3$ at room temperature have been predicted using theoretical group analysis [31]. The $A_1$ and E modes are both Raman and IR-active modes, whereas the $A_2$ modes are Raman and IR inactive modes. Where, $A_1$ modes are polarized along z-axis and E modes in x-y plane. By fitting the measured spectra and deconvoluting the fitted curves into individual lorentzian components, the peak position of each component, i.e. the natural frequency ($cm^{-1}$) of each Raman active mode, has been obtained for all the ceramics and the typical fitting for BFO has been shown in Fig. 5 (a). Two sharp peaks at around 143 and 175 $cm^{-1}$ for BFO merged and results in the peak at 158 $cm^{-1}$ for the co-substituted samples which could be assigned as $A_1$-1 and Peak at around 213 $cm^{-1}$ as $A_1$-3 phonon modes respectively. E modes are assigned to other phonon modes located in the range of 50 – 750 $cm^{-1}$ as shown in Fig. 5 (a) & (b). The frequency of the mode is proportional to $(k/M)^{1/2}$, where, k is the force constant and M is the reduced mass. The average mass of the A-sites significantly decreases with the increase in substitution concentration because atomic mass of La is about 66% of that of Bi which leads to shift of $A_1$ phonon modes to higher frequency side as the frequency of the phonon mode is inversely proportional to the reduced mass. It suggests the chemical pressure-induced bond shortening and lattice distortion [32]. Based on the previous literature, these results demonstrate the existence of the chemical pressure-induced rhombohedral–orthorhombic phase transition in BFO by co-substitution of La & Mn in Bi & Fe sites respectively. Similar results about the phase transition from the rhombohedral to orthorhombic symmetry in bismuth ferrite under high pressure using Mao-Bell diamond anvil cell have been observed by Y. Yang et al [33]. However, the Raman spectra of the samples with co-substitution concentration above 15 % shows the strong band at around 620 $cm^{-1}$ that is similar to orthorhombic $RMnO_3$ (R = rare earth) as shown in Fig. 5 [34]. It clearly suggests the increase in the phase percentage of orthorhombic crystal symmetry with the increase in the substitution percentage.

Figure 6 shows the FT-IR spectra for present ceramics in the wave number range of 400-750 $cm^{-1}$. Typical band characteristics of metal oxygen bonds were observed in the range of 425-575 $cm^{-1}$. The absorption peaks at around 553 and 452 $cm^{-1}$ are due to Fe-O bond stretching and O-Fe-O bending of $FeO_6$ group in the perovskite structure. The absorption peak around 669 $cm^{-1}$ corresponds to the water vapor. These band positions are found to be in agreement with the characteristic infrared absorption bands of BFO [35]. With the increase in substituents percentage the absorption peaks should shift to the higher wave number side due to increased lattice strain resulting from the difference in ionic radius of substituent and host cations. With the increase in substituents percentage above 15 %, the transmission bands shift to the lower wave number because the lattice strain relaxes with the dominance of orthorhombic crystal symmetry fraction.

In order to find out the effect of chemical pressure on the band gap of BFO, we have recorded the diffuse reflectance spectra of all the ceramics. UV-Vis diffuse reflectance spectra were converted into absorption readings according to the Kubelka-Munk (K-M) method [36]. The absorption spectrum of the ceramics transformed from the diffuse reflection spectra using Kubelka-Munk function,

$$F(R) = \frac{(1-R)^2}{2R} \qquad \text{- - - - - - - - - - - - - (4)}$$

Where, R is diffuse reflectance. Since BFO has a distorted cubic perovskite structure, there is a point group symmetry breaking from $O_h$ to $C_{3v}$ [37, 38]. There are expected six transitions between 0 and 3 eV by considering $C_{3v}$ local symmetry of $Fe^{3+}$ ions (High spin configuration $t_{2g}^3 e_g^2$ ) in BFO and using the correlation group and subgroup analysis for the symmetry breaking from $O_h$ to $C_{3v}$ [39]. In our case all six transitions were observed which lie in the range between 1.3 to 3 eV for BFO as shown in Fig. 7 (a). However, these peaks vanish with the increase in the substitution concentration and have low relative intensity in BLFM-05 (Fig. 7 (b)). This indicates the structural transition from the distorted cubic perovskite structure to orthorhombic symmetry due to internal chemical pressure which arises as a result of size mismatch between substitution and host cations which corresponds to the modification in $FeO_6$ local environment. The analysis of XRD patterns, Raman spectra, FTIR spectra as discussed above also support well in this context.

The magnetization versus magnetic field (M-H) plots at room temperature of all ceramics with a maximum applied magnetic field of ± 80 kOe have been shown in Fig. 8 and magnetic parameters have been enlisted in table 4. The magnetic moment of $Fe^{3+}$ cations in BFO , are ferromagnetically coupled in pseudo cubic (111) planes but antiferromagnetically between adjacent planes and it is surrounded by six $O^{2-}$ ions in the common vertex of two adjacent $FeO_6$ octahedra. Magnetic hysteresis loop shows the linear magnetic field dependence of magnetization for BFO which indicates that it is antiferromagnetic material. However, the co-substitution of La and Mn induces weak ferromagnetism. The unsaturated hysteresis loops and presence of small remnant magnetization reveals the presence of antiferromagnetic with weak ferromagnetism.

The tolerance factor of the substituted ceramics decreases with the increase in concentration of substituted elements because ionic size of $La^{3+}$ is smaller than that of $Bi^{3+}$. This leads to the increase in the octahedral tilt and the Fe-O-Fe bond angles which, in turn increases the canting angle of the antiferromagnetically ordered adjacent planes with the suppression of cycloid spin structure [40]. The evolution of weak ferromagnetism in substituted ceramics is accounted due to the canting of antiferromagnetically ordered spins because of structural distortion as the there is no contribution of magnetization from the impurity phase as $Bi_2Fe_4O_9$ and $Bi_{25}FeO_{40}$ because it is paramagnetic at room temperature [41]. The maximum magnetization increases with increasing the substitution concentration due to gradual increase in local lattice distortion. The remnant magnetization increases with the increase in the substitution concentration and reaches a maximum at x = 0.050, which indicates almost broken spin cycloid structure. The further increase in substitution concentration results in the reduction of remnant magnetization due to the appearance of complete antiferromagnetic ordering in the orthorhombic structure because of the significant contribution from the crystallographic phase of *Pbnm* space group (as obtained from the quantitative crystallographic phase contribution by double phase Rietveld analysis).

The substitution and other effects can break the cycloids in BFO, which has been reported to affect the magnetic properties of the ceramics, but presence of magnetic impurities ($Fe_2O_3$ or $Fe_3O_4$) may be the cause for the same. In our case, the presence of $Fe_2O_3$ magnetic impurity has not been detected form Rietveld analysis of XRD patterns. Moreover, we have not observed ferromagnetic hysteresis loop for BFO which has largest amount of impurity phases among all ceramics. Also, the area of ferromagnetic hysteresis loop and coercivity increases with the increase in substitution percentage upto 5 % and then decreases due to appearance of complete antiferromagnetic ordering in the orthorhombic structure. It also support the crystallographic phase percentage obtained from Rietveld analysis of XRD patterns which shows that the percentage of the orthorhombic crystal symmetry (*Pbnm* space group) is significant in the 5 % of co-substitution. This result can be attributed to the fact that the weak ferromagnetism in the substituted ceramics might be due to broken cycloid spin structure which leads to canting of the antiferromagnetic spin structure.

The limited device applications of BFO are due to high leakage current, which can be reduced by compensating the oxygen vacancies using appropriate substituents. Hence, the effect of co-substitution by La and Mn in BFO on the room temperature dielectric constant as well as loss versus frequency measurements of all ceramics has been studied in the frequency range of 500 Hz – 5 MHz (shown in Fig. 9 (a) & (b)). The dielectric constant decreases rapidly with the frequency in lower frequency range ($< 10^4$ Hz) which is due to significant oxygen vacancies and secondary phases. It is independent in higher frequency range ($> 10^5$ Hz) because dipoles with large effective masses (e.g., oxygen vacancies, segregated impurities phases) are able to respond at low frequency of the applied field but, not at higher frequencies. The decrease of dielectric constant with the increase in frequency can be explained by dipole relaxation phenomenon which is due to Maxwell – Wagner type of interfacial polarization contribution.

The low frequency dielectric constant (Fig. 9(a)) increases with the co-substitution percentage which suggests that although the impurity phases decrease with the substitution, but oxygen vacancies may be increasing which has been observed from EDS analysis. The dielectric constant at high frequency region ($> 10^5$ Hz) increases with the increase in substitution percentage and attains maximum for BLFM-05 ceramics, then decreases for the higher substitution. At the same time, dielectric loss for this particular composition was observed to be minimum which could be due to reduction of impurity phase and smaller grain size for the substituted BFO which results in the increase of grain boundaries acting as scattering centre for the moving electrons.

**4. Conclusions**

The co-substitution of La and Mn has resulted in structural transition from rhombohedral symmetry (space group *R3c*) to orthorhombic (*Pbnm*) as indicated by the XRD as well as Raman spectra analysis. The crystallographic phase contribution has been quantified by double phase Rietveld analysis of XRD patterns. It shows that the orthorhombic crystal symmetry (*Pbnm* space group) is significant beyond 5% of co-substitution which results in the sharp decrease of the $M_r$ and $H_C$ in this case due to appearance of complete antiferromagnetic ordering (in the orthorhombic crystal structure). The ceramics with 5% of co-substitution have enhanced remnant magnetization about 15 times that of pure one due to the suppression of cycloid spin structure and enhanced canting angle of antiferromagnetically ordered spins caused by the crystal lattice distortion. The improved dielectric properties with very low value of dielectric loss have been observed for the BLFM-05. Raman, FTIR and UV-Visible spectra analysis also supports the modification in local $FeO_6$ environment and structural transition. The enhanced dielectric and magnetic properties may be interesting for device applications.

**Acknowledgment**

The authors gratefully acknowledge Dr. Dhanvir Singh Rana and his research group at IISER Bhopal for extending the MPMS facility.

**Table 1** Structural parameters of $Bi_{1-x}La_xFe_{1-x}Mn_xO_3$ (for x = 0.000 - 0.300) ceramics annealed at 700 $^oC$. Where, x = substitution value, D = crystallite size, t = tolerance factor, $\chi 2$ = Chi-square (goodness of fit).

| **X** | **D** | **Fe-O-Fe Angle (*R3c*)** | **V (Å$^3$) (*R3c*)** | **t** | **$\chi$2** |
|---|---|---|---|---|---|
| 0.000 | 82.49 | 154.487 | 372.470 | 0.8888 | 1.49 |
| 0.015 | 76.52 | 154.820 | 371.050 | 0.8886 | 1.47 |
| 0.025 | 50.78 | 155.215 | 370.034 | 0.8885 | 1.42 |
| 0.050 | 45.90 | 156.626 | 369.003 | 0.8884 | 1.31 |
| 0.100 | 33.50 | 161.906 | 367.140 | 0.8883 | 1.37 |
| 0.150 | 31.02 | 167.402 | 363.430 | 0.8881 | 1.22 |
| 0.200 | 35.82 | 166.852 | 370..851 | 0.8879 | 1.34 |
| 0.250 | 37.32 | 165.220 | 371.004 | 0.8877 | 1.22 |
| 0.300 | 39.14 | 160.001 | 372..410 | 0.8875 | 1.30 |

**Table 2:** Fractional atomic coordinates for the Rietveld refinement of XRD patterns of $Bi_{0.950}La_{0.050}Fe_{0.950}Mn_{0.050}O_3$.

| **Space Group** | ***R3c*** | ***Pbnm*** |
|---|---|---|
| Bi/La | 6a (0, 0, 0) | 4c (0.9976(7), 0.0232(5), 0.2500(3)) |
| Fe/Mn | 6a (0, 0, 0.2172(5) | 4b (0.5000(2), 0.0000(5), 0.0000(1)) |
| O(1) | 18b (0.4579(3), 0.0264(7) 0.9497(0)) | 4c (0.0661(4), 0.4799(7), 0.2500(3)) |
| O(2) | | 8d (0.7271(5), 0.2826(4), 0.0313(7)) |

**Table 3:** Crystallographic Phase contribution of all ceramics obtained by the Rietveld refinement of XRD patterns for $Bi_{1-x}La_xFe_{1-x}Mn_xO_3$ for x = 0.000 - 0.300.

| | **Crystallographic Phase contribution of all ceramics** | |
|---|---|---|
| $Bi_{1-x}La_xFe_{1-x}Mn_xO_3$ | ***R3c*** | ***Pbnm*** |
| X = 0.000 | 100 % | 0 % |
| X = 0.015 | 93.14 % | 6.86 % |
| X = 0.025 | 92.96 % | 8.04 % |
| X = 0.050 | 87.37 % | 12.63 % |
| X = 0.075 | 82.48 % | 17.52 % |
| X = 0.100 | 75.61 % | 24.39 % |
| X = 0.150 | 49.41 % | 50.59 % |
| X = 0.200 | 43.65 % | 56.35 % |
| X = 0.250 | 40.53 % | 59.47 % |
| X = 0.300 | 38.06 % | 61.94 % |

**Table 4:** Magnetic parameters of $Bi_{1-x}La_xFe_{1-x}Mn_xO_3$ (for x = 0.000 - 0.200) ceramics annealed at 700 $^oC$. Where, $M_S$ = magnetization at maximum applied field, $H_C$ = Coercive field, emu/g = emu/gram at applied magnetic field of 80 kOe.

| $Bi_{1-x}La_xFe_{1-x}Mn_xO_3$ | $M_S$ at 80 kOe (emu/g) | $M_r$ (emu/g) | $H_C$ (kOe) |
|---|---|---|---|
| X = 0.000 | 0.6420 | 0.0086 | 0.1692 |
| X = 0.015 | 0.7827 | 0.0514 | 5.8524 |
| X = 0.025 | 0.8312 | 0.0916 | 9.0252 |
| X = 0.050 | 0.8598 | 0.1282 | 9.3701 |
| X = 0.100 | 0.8635 | 0.0554 | 4.8739 |
| X = 0.150 | 1.0105 | 0.0311 | 1.8006 |
| X = 0.200 | 1.0922 | 0.0128 | 0.7475 |

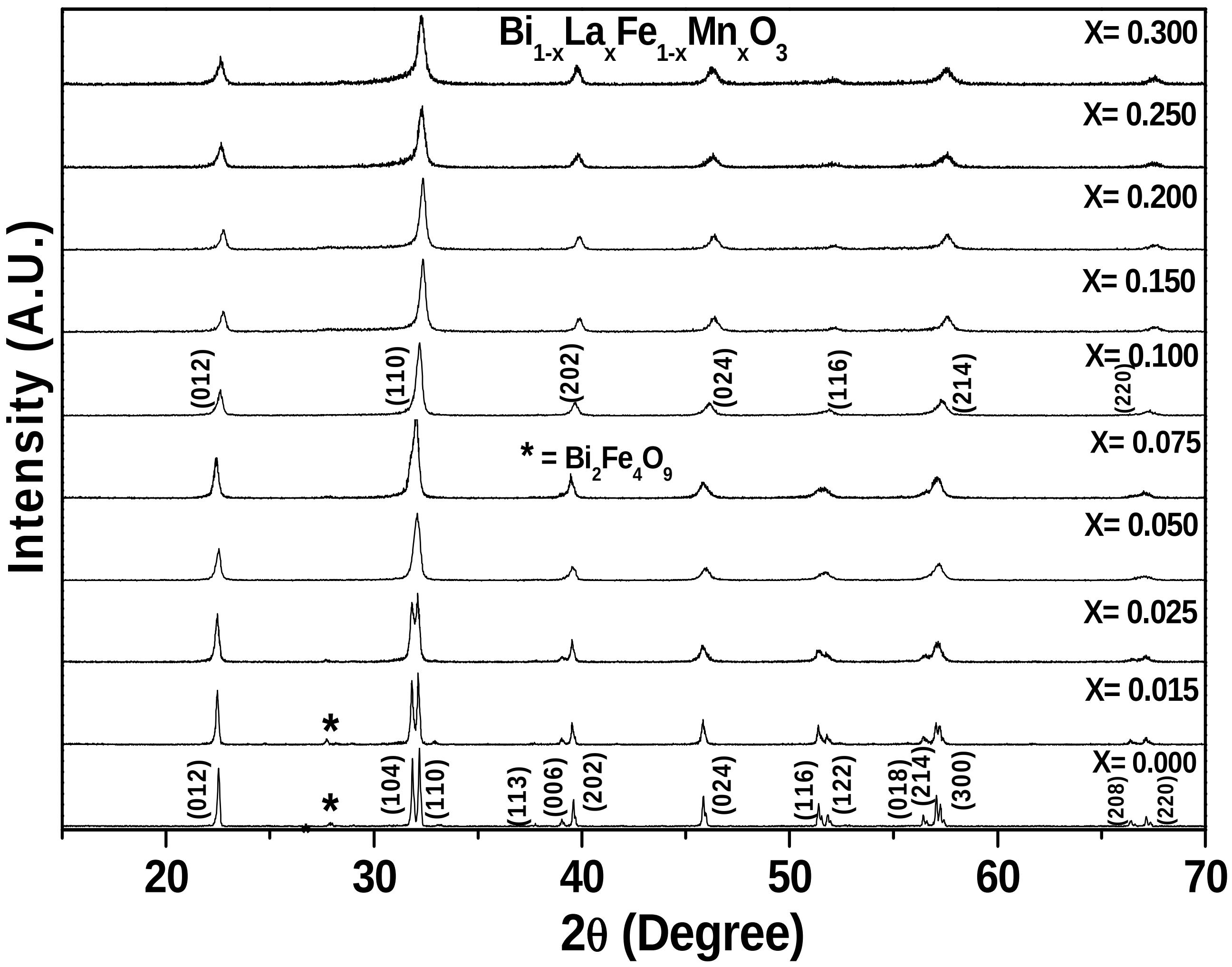

**Fig. 1.** XRD patterns of $Bi_{1-x}La_xFe_{1-x}Mn_xO_3$ ceramics for x = 0.000 – 0.300.

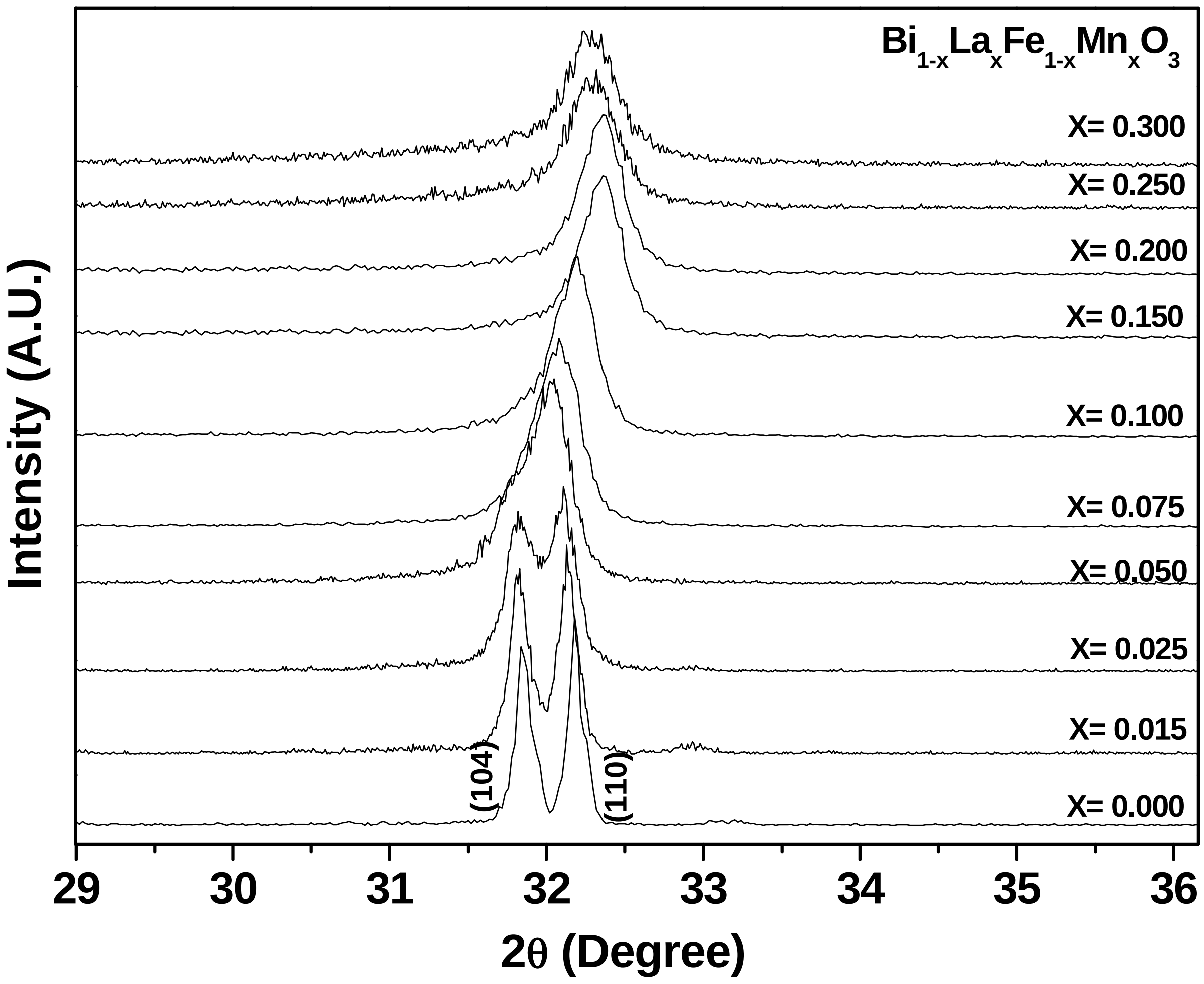

**Fig. 2.** Shift in the XRD patterns of $Bi_{1-x}La_xFe_{1-x}Mn_xO_3$ ceramics for x = 0.000 – 0.300.

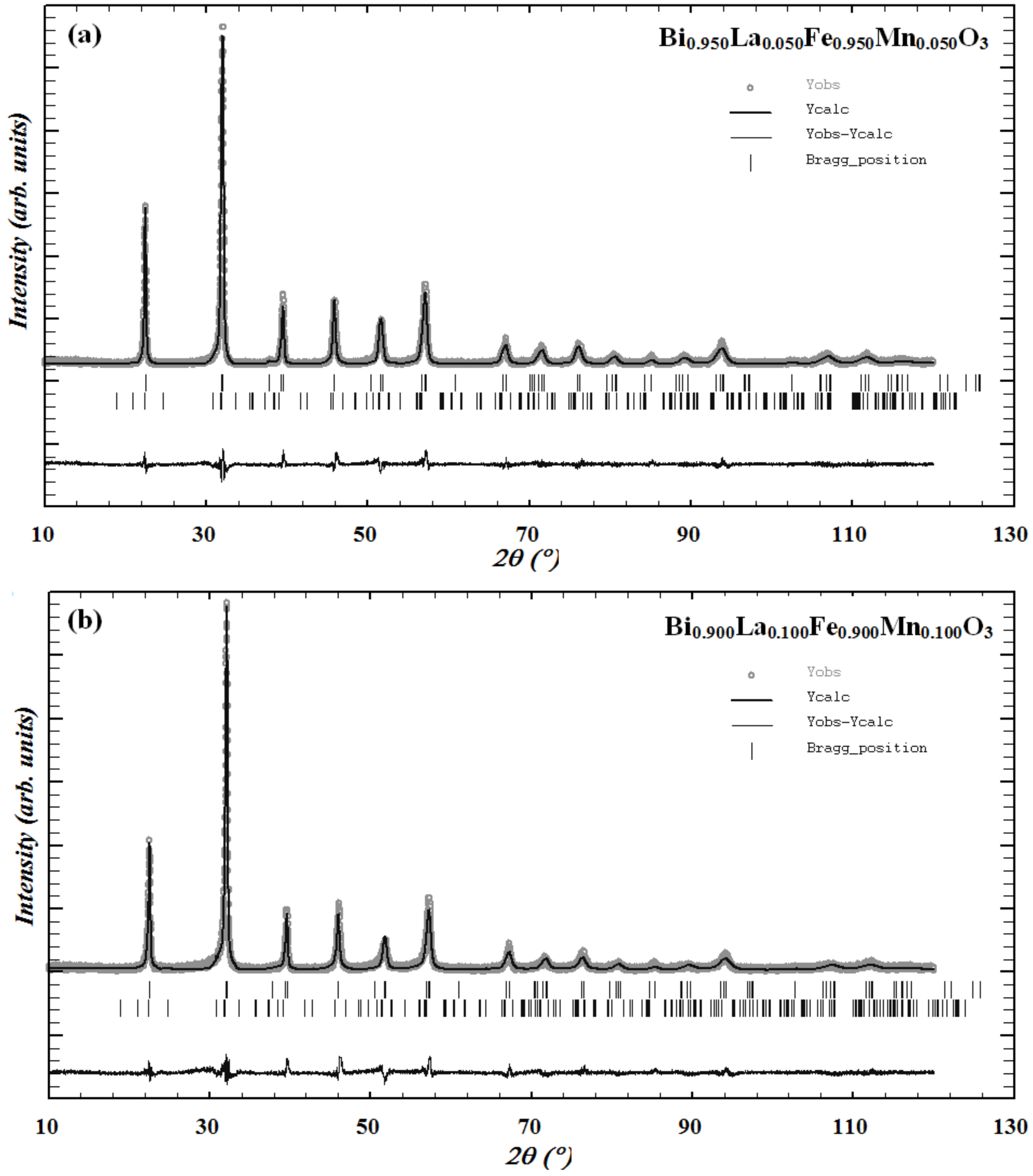

**Fig. 3.** Rietveld refined XRD patterns of $Bi_{1-x}La_xFe_{1-x}Mn_xO_3$ ceramics for (a) x = 0.050 and (b) x = 0.100.

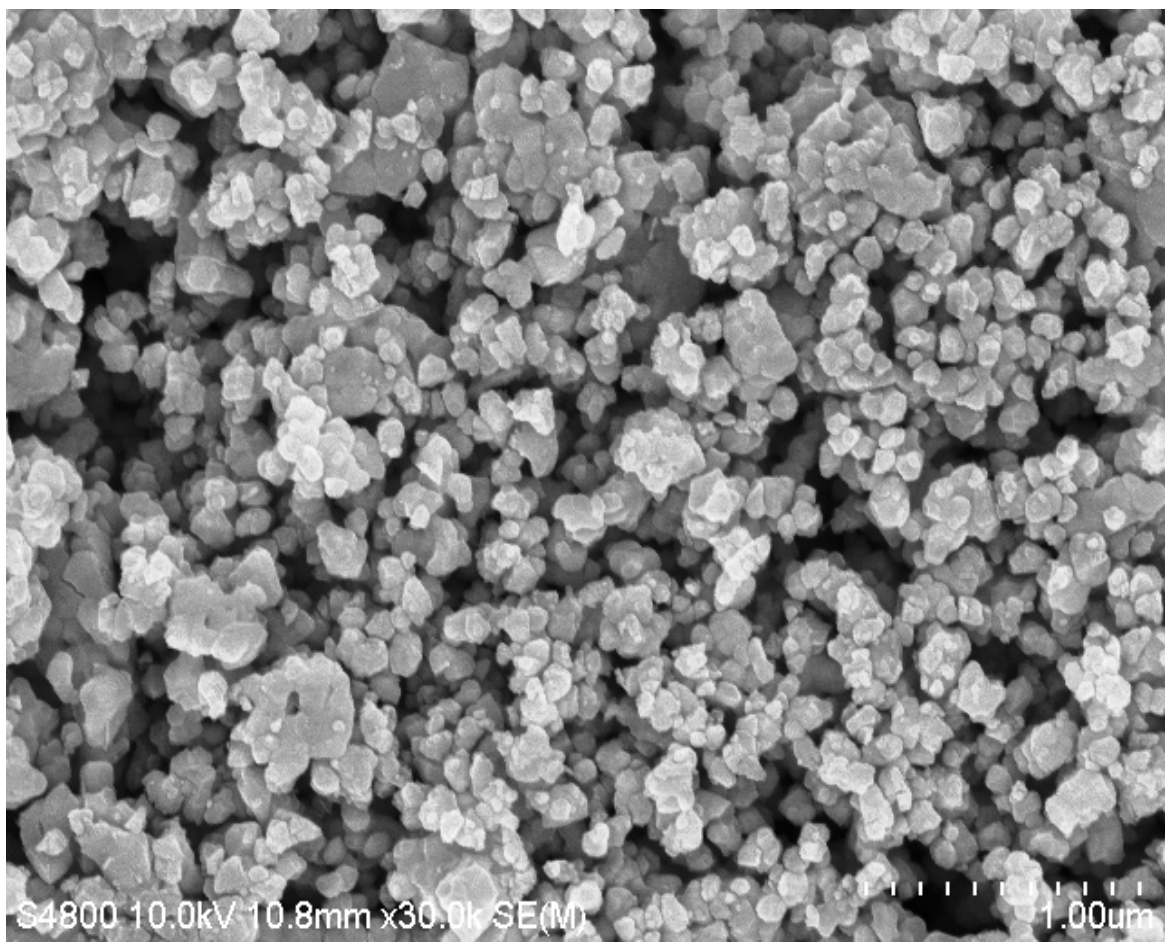

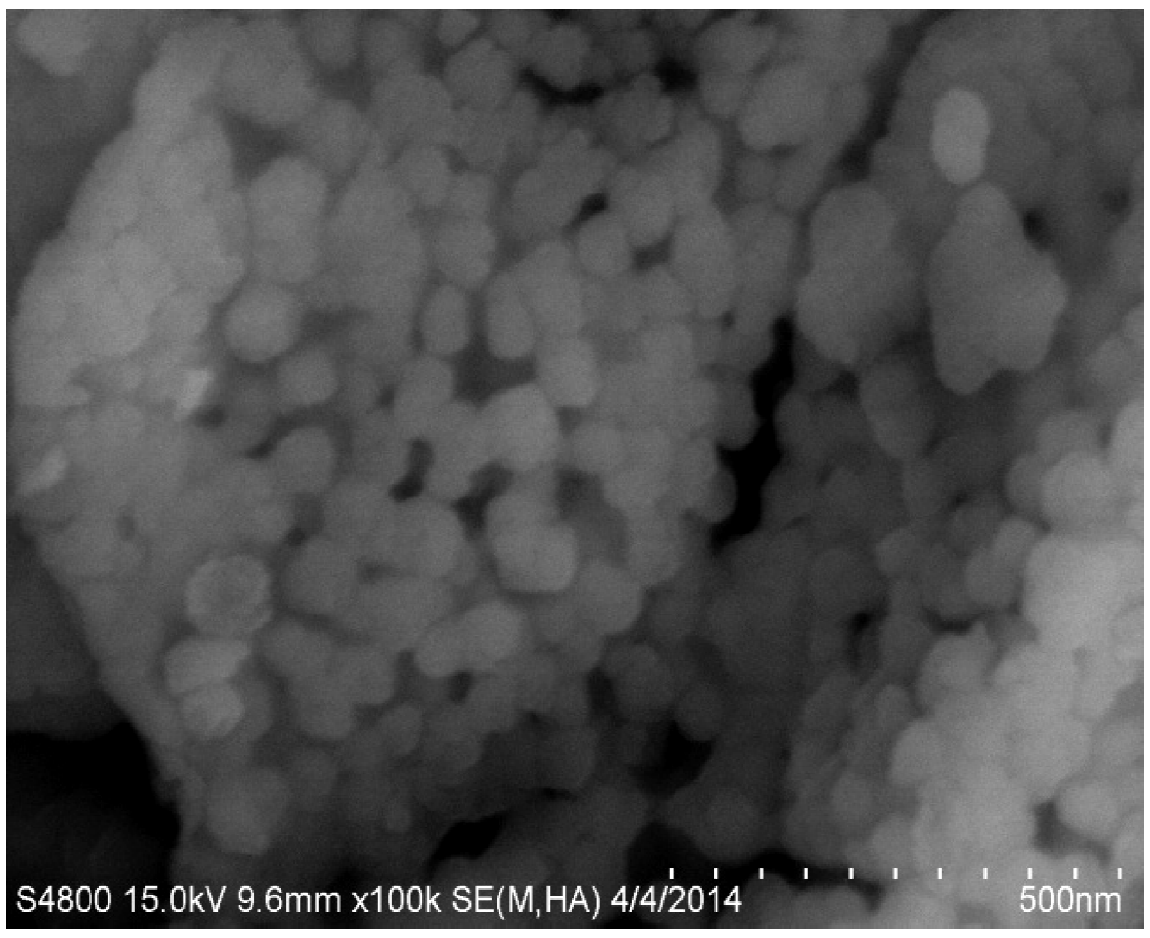

(a) (b)

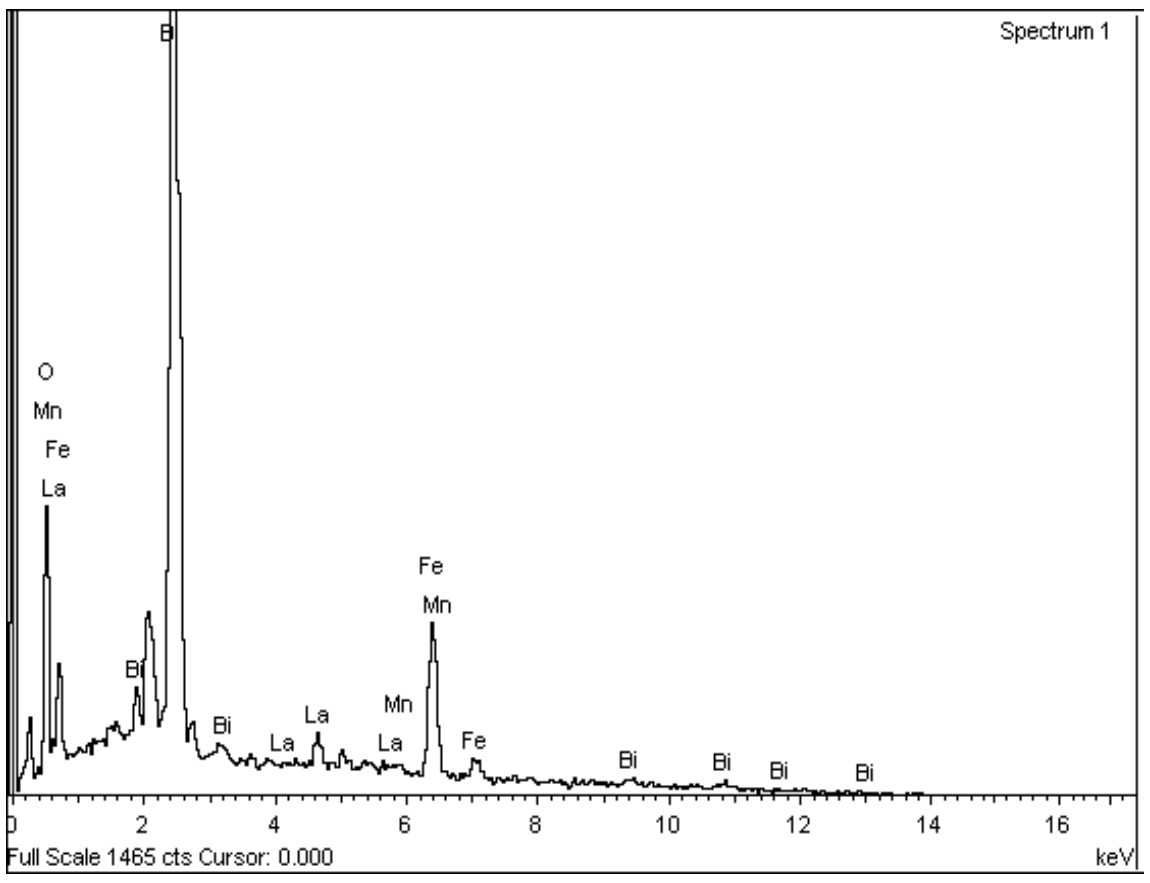

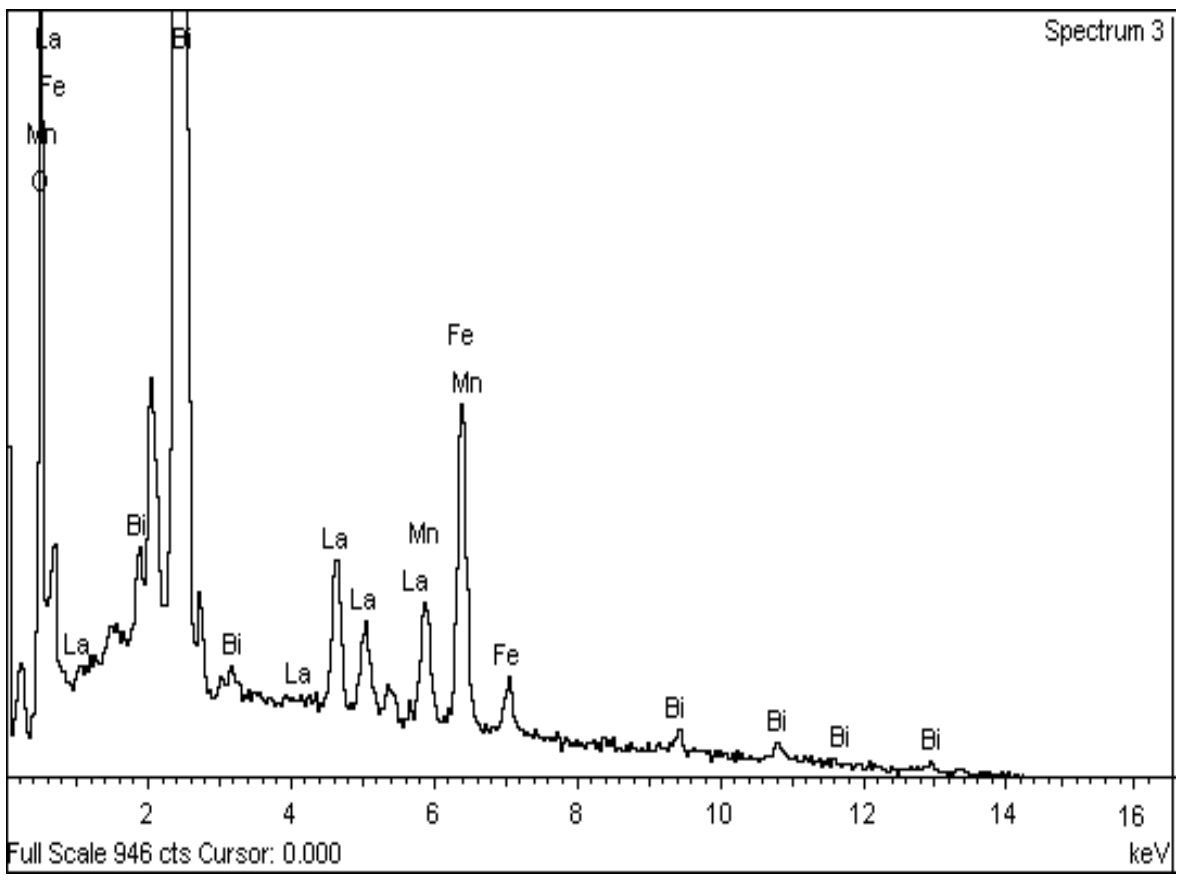

(c) (d)

**Fig. 4**. FE-SEM image and EDS pattern of $Bi_{1-x}La_xFe_{1-x}Mn_xO_3$ ceramics for x = 0.025 & x = 0.050 respectively in figures (a), (b), (c) and (d).

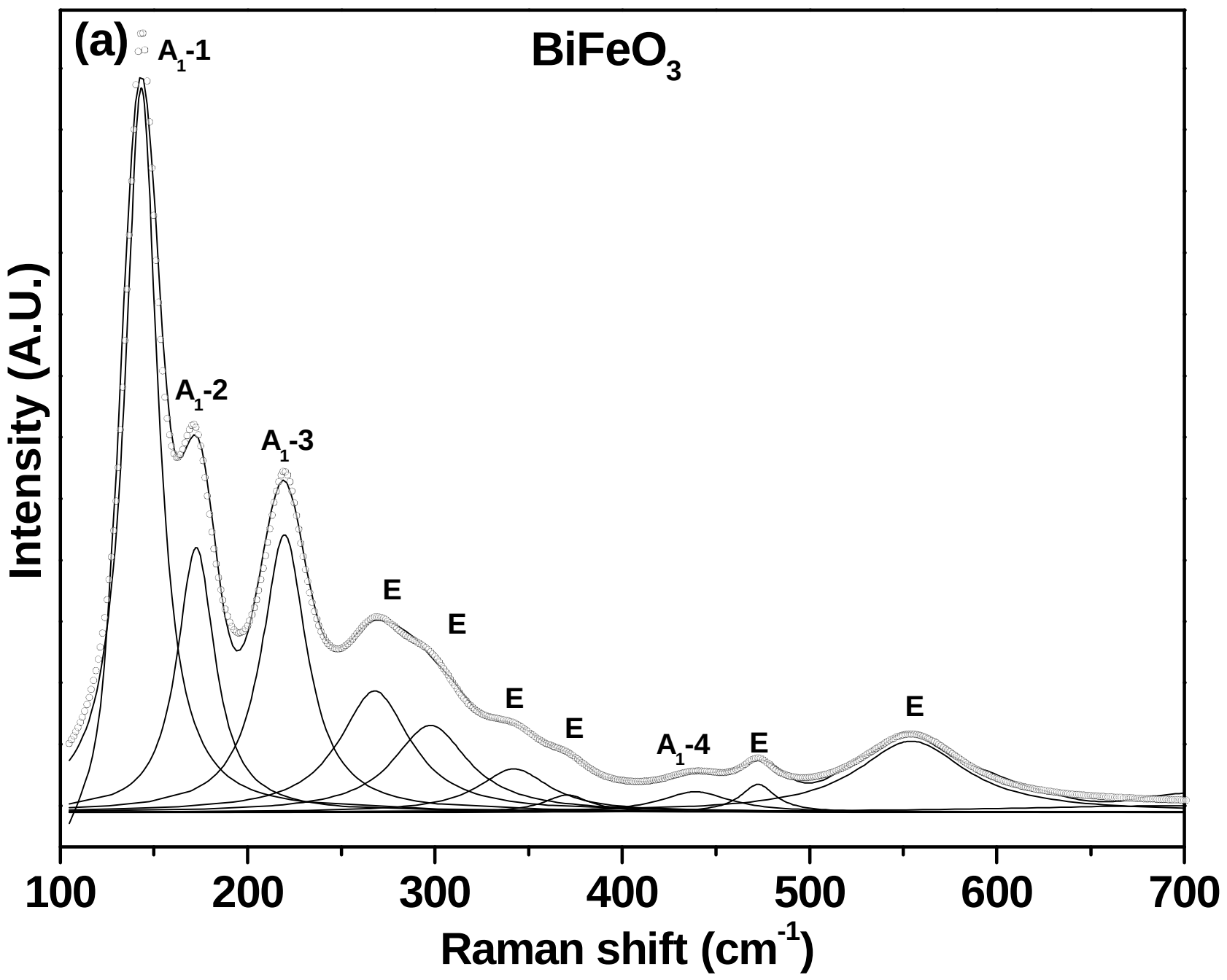

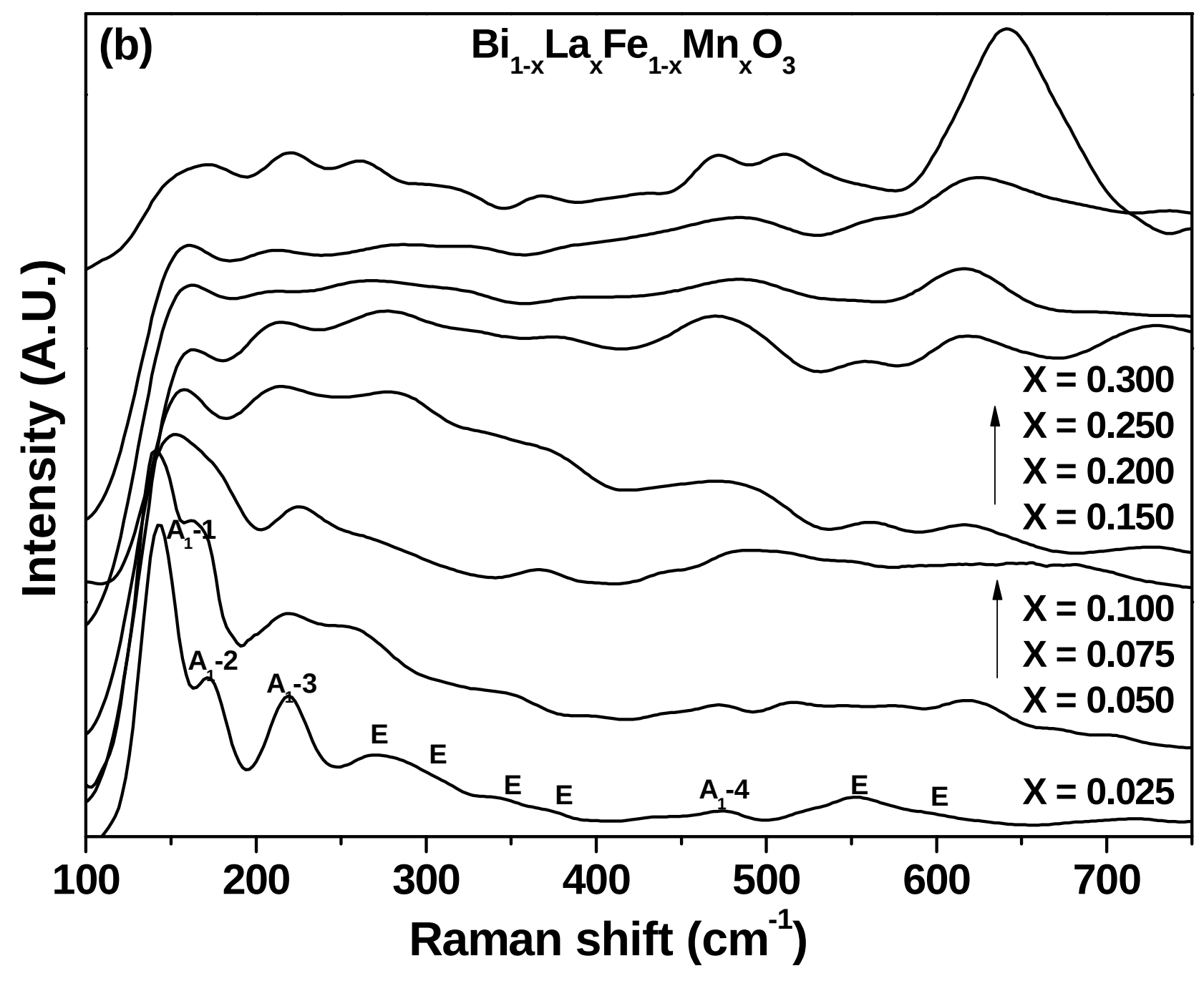

**Fig. 5.** Typical (a) Raman scattering spectra of $BiFeO_3$ with its phonon modes deconvoluted into individual lorentzian components and (b) combined spectra of $Bi_{1-x}La_xFe_{1-x}Mn_xO_3$ ceramics for x = 0.025 - 0.300.

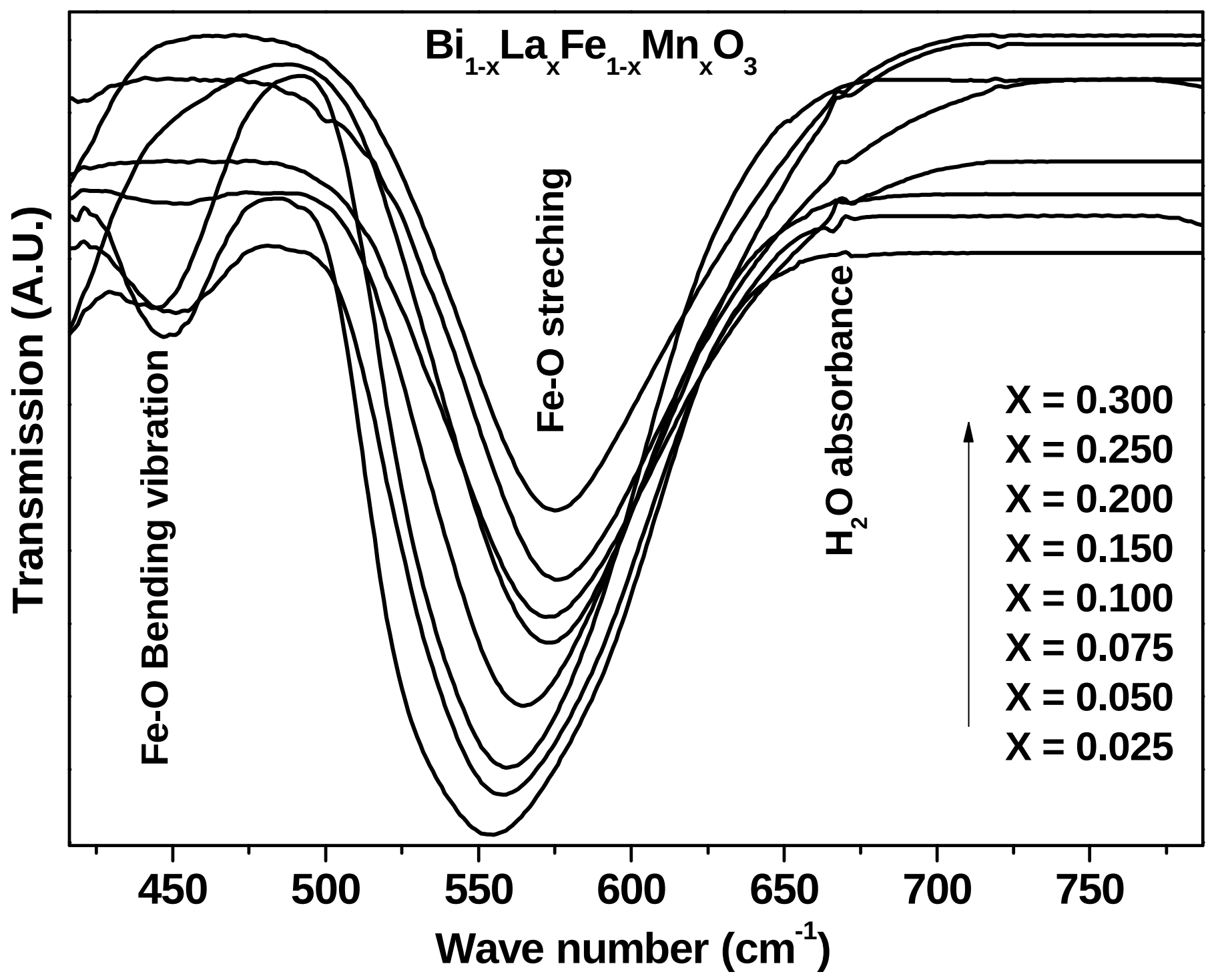

**Fig. 6.** FT-IR spectrum of $Bi_{1-x}La_xFe_{1-x}Mn_xO_3$ ceramics for x = 0.025 - 0.300.

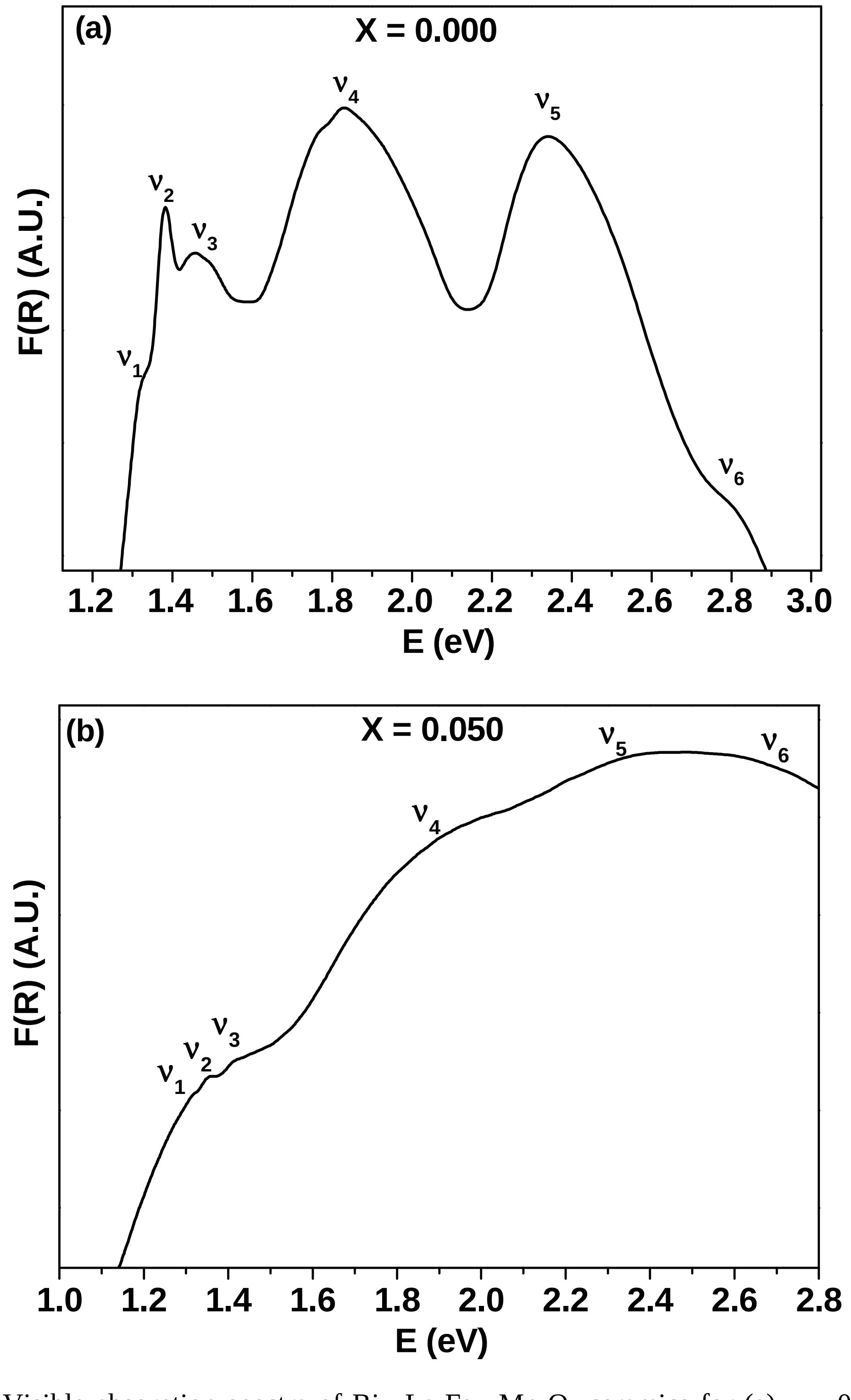

**Fig. 7**. UV-Visible absorption spectra of $Bi_{1-x}La_xFe_{1-x}Mn_xO_3$ ceramics for (a) x = 0.000 & (b) 0.050 in the 1 – 3.0 eV energy range.

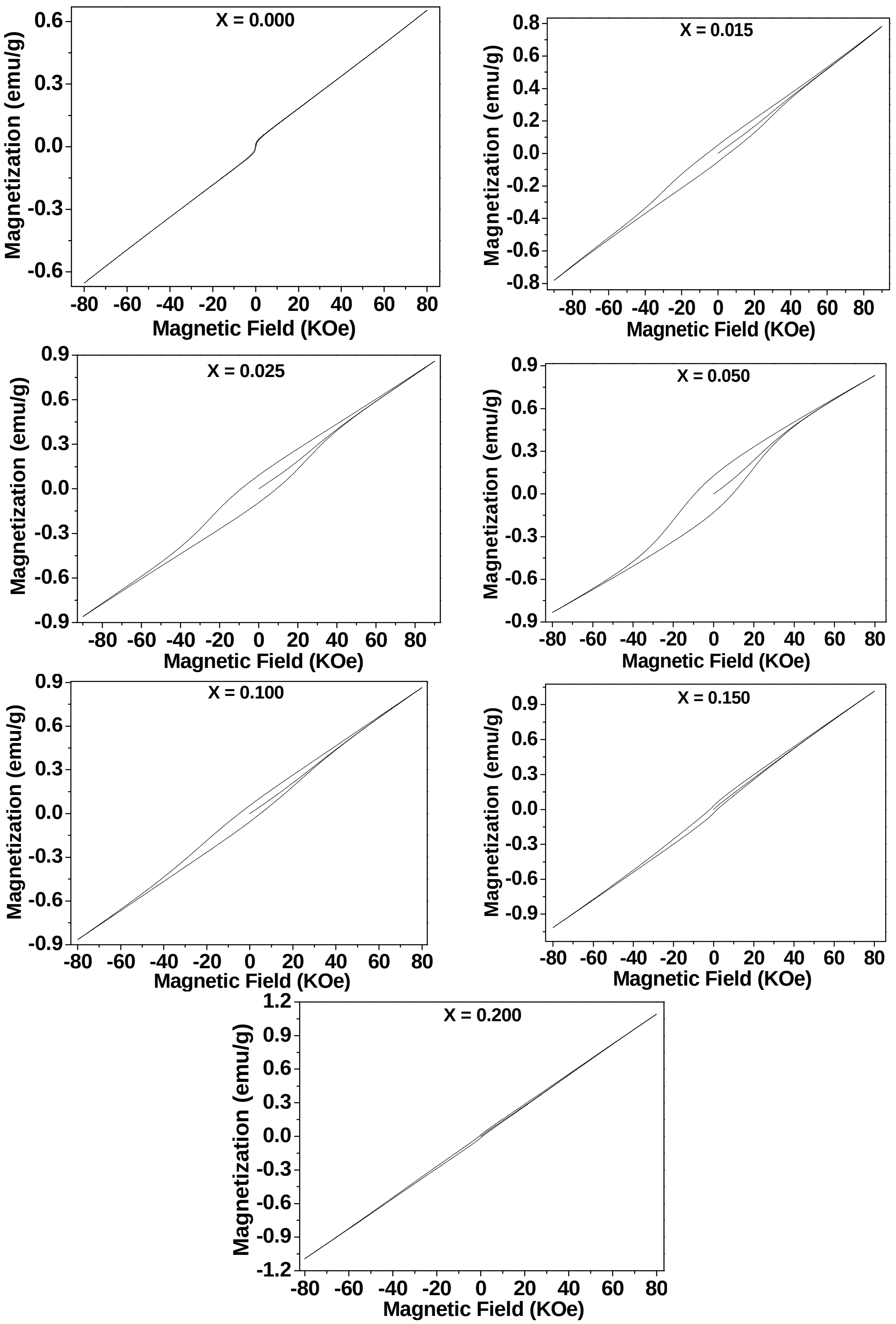

**Fig. 8**. M-H loops of $Bi_{1-x}La_xFe_{1-x}Mn_xO_3$ ceramics for x = 0.000 - 0.200.

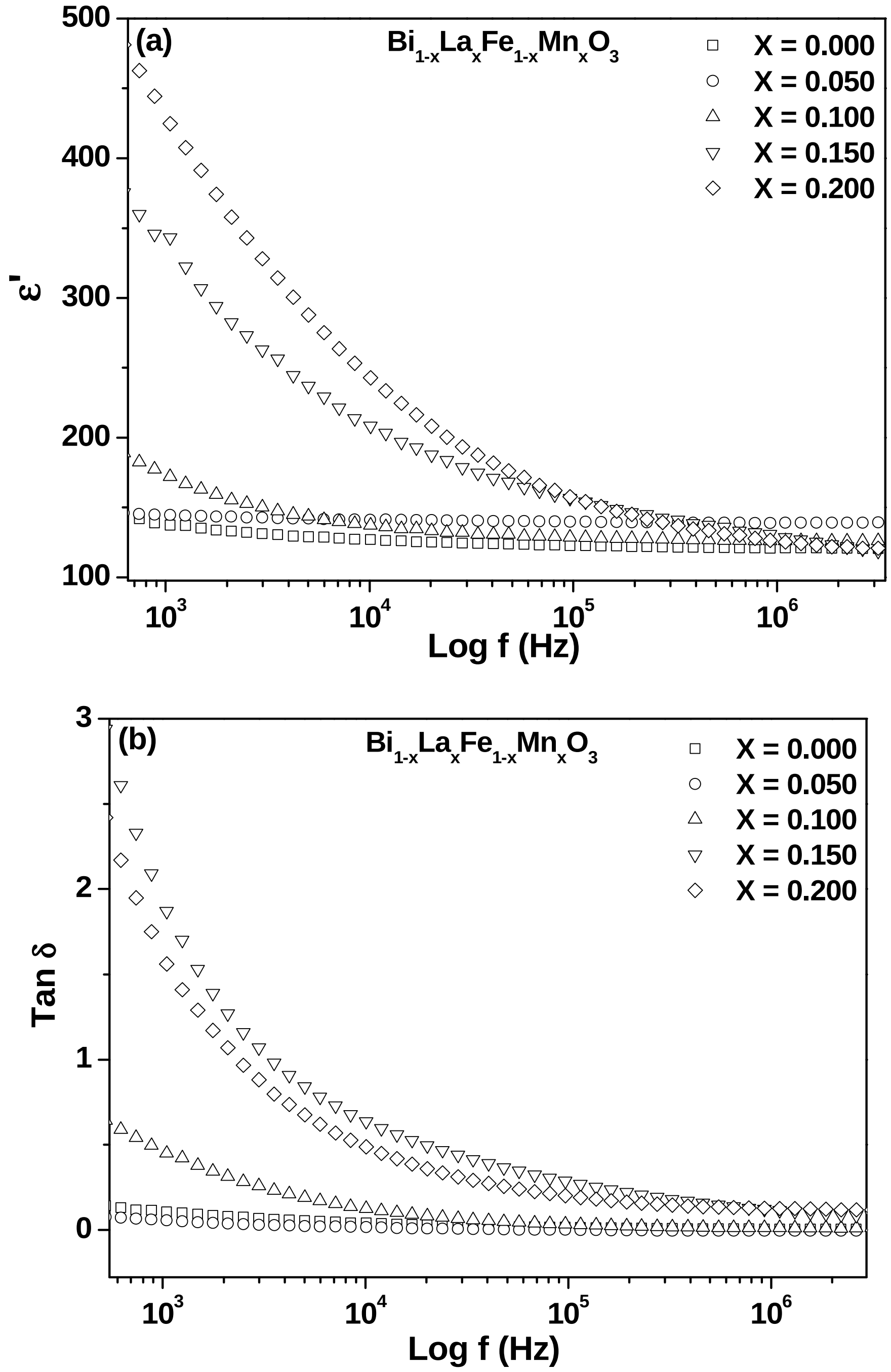

**Fig. 9**. Room temperature (a) dielectric constant (b) dielectric loss versus frequency (plotted at log scale) plot of $Bi_{1-x}La_xFe_{1-x}Mn_xO_3$ for x = 0.000 - 0.200.